\documentclass[a4paper,UKenglish,cleveref,autoref, thm-restate,authorcolumns,notab,nolineno]{socg-lipics-v2021}

\hideLIPIcs

\usepackage{xspace}

\newcommand{\cR}{{\cal R}}

\newcommand{\ZZ}{{\mathbb Z}}

\def\b1{{\bf 1}}

\usepackage{svg}
\usepackage{subcaption}

\graphicspath{{./graphics/}}

\title{Coordinated Motion Planning Through Randomized $k$-Opt}

\titlerunning{CG:SHOP Challenge 2021}

\author{Paul Liu}{Department of Computer Science, Stanford University, USA \and \url{https://cs.stanford.edu/people/paulliu/}}{paul.liu@stanford.edu}{https://orcid.org/0000-0002-9386-6609}{}

\author{Jack Spalding-Jamieson
}{Department of Computer Science, University of Waterloo, Canada}{jacksj@uwaterloo.ca}{}{}

\author{Brandon Zhang
}{No affiliation}{brandon.zhang@alumni.ubc.ca}{}{}

\author{Da Wei Zheng}{Department of Computer Science, University of Illinois at Urbana-Champaign, USA \and \url{https://davidzheng.web.illinois.edu/}}{dwzheng2@illinois.edu}{https://orcid.org/0000-0002-0844-9457}{}

\authorrunning{Paul Liu, Jack Spalding-Jamieson, Brandon Zhang,  Da Wei Zheng} 

\Copyright{Paul Liu, Jack Spalding-Jamieson, Brandon Zhang, Da Wei Zheng} 


 % TODO

\category{CG challenge}

\supplement{github.com/jacketsj/cgshop2021-gitastrophe}

\EventEditors{Kevin Buchin and \'{E}ric Colin de Verdi\`{e}re}
\EventNoEds{2}
\EventLongTitle{37th International Symposium on Computational Geometry (SoCG 2021)}
\EventShortTitle{SoCG 2021}
\EventAcronym{SoCG}
\EventYear{2021}
\EventDate{June 7--11, 2021}
\EventLocation{Buffalo, NY, USA}
\EventLogo{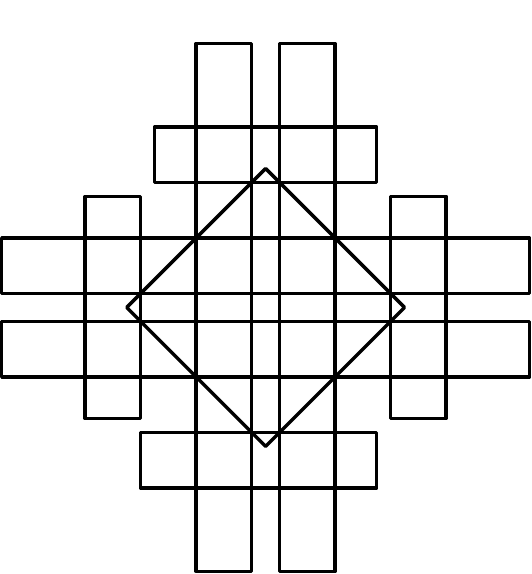}
\SeriesVolume{189}
\ArticleNo{64}
\ArticleNo{83}

\begin{document}

\maketitle

\everypar{\looseness=-1}
\linepenalty=20000

\begin{abstract}
This paper examines the approach taken by team {\tt gitastrophe} in the CG:SHOP 2021 challenge. 
The challenge was to find a sequence of simultaneous moves of square robots between two given configurations that minimized either total distance travelled or makespan (total time).
Our winning approach has two main components: an initialization phase that finds a good initial solution, and a $k$-opt local search phase which optimizes this solution. This led to a first place finish in the distance category and a third place finish in the makespan category.
\end{abstract}

\section{Introduction}
For a set of unit square robots $\cR$ each with start and target locations $\{s_r\}_{r\in \cR}, \{d_r\}_{r\in \cR} \subset \ZZ^2$, and a set of unit square obstacles $O\subset \ZZ^2$, the \emph{coordinated motion planning} problem asks for an optimal sequence of simultaneous ``moves'' for each robot that brings the robots from their start location to their target location. At each timestep, robots can move to an adjacent grid cell or stand still (a no-op). These moves are subject to the following constraints at all timesteps:
\begin{itemize}
    \item No robot shares a location with another robot or an obstacle.
    \item No robot moves into a location previously occupied by another robot, unless the other robot is moving in the same direction at that timestep.
\end{itemize}
An \emph{optimal} sequence of moves can refer to two different criteria:
\begin{itemize}
    \item \textbf{MAX:} The minimum {\bf makespan}, i.e. the total number of timesteps needed.
    \item \textbf{SUM:} The minimum {\bf total distance}, i.e. the total number of position-changing moves each robot takes.
\end{itemize}

We, team {\tt gitastrophe},
explored this problem as a part of the 2021 Computational Geometry Challenge (CG:SHOP 2021). The challenge ranked teams according to each of the two different optimality criteria.
Our team ranked first among all junior teams, first according to the total distance criterion, and third according to the makespan criterion. Team Shadoks \cite{shadoks} ranked first according to makespan and third according to total distance, while Team UNIST \cite{unist} ranked second in both categories.

One important property of the test instances of the challenge is that feasibility is always guaranteed --- no obstacles enclose the starting or target positions of any robots. In the following sections, we assume this property. We refer the reader to the survey of the challenge~\cite{cgchallenge} for more details on the instances.

\section{Methods}
As with other teams who took part in the challenge, our strategy consisted of two phases: initialization (\Cref{sec:init}) and solution optimization (\Cref{sec:k-opt}). 

The initialization phase consisted of moving the robots out of the initial problem grid, into an intermediate state that can easily be routed to their end goals. This strategy was remarkably similar to the feasibility approaches used by both Team Shadoks \cite{shadoks} and Team UNIST \cite{unist}.
The solution optimization phase used a novel local search strategy, which consisted of locally reordering the movements of $k$ sampled robots while keeping all others fixed.

\subsection{Initialization strategy}
\label{sec:init}
Our initialization strategy is simple and relies on the notion of \emph{depth} values (see \Cref{def:depth}). The depth values partition the robots into subsets which can be routed in parallel. Intuitively, once a robot is routed, it will not interfere with robots of smaller depth value.

\begin{figure}[ht]
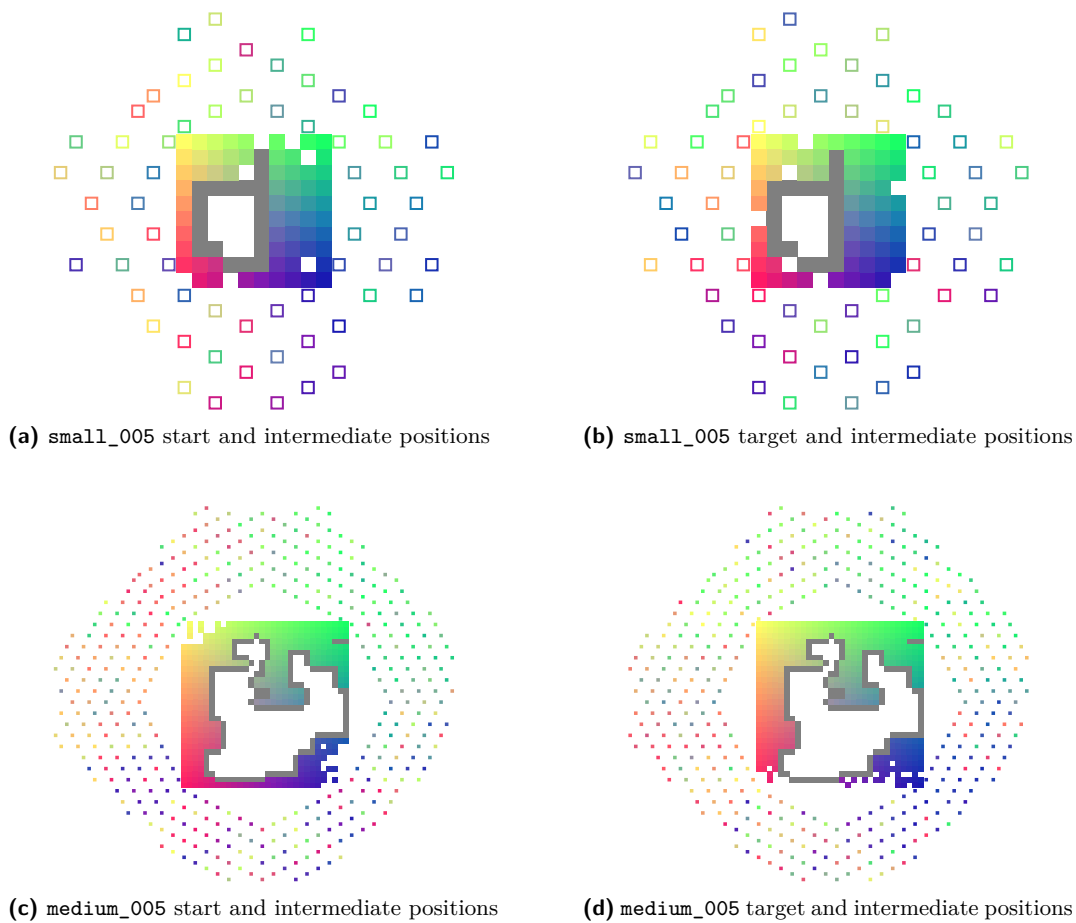

    \centering
    \begin{minipage}[t]{.46\textwidth}
        \centering
        \includesvg[scale=1.0]{graphics/ff1_small_005_10x10_90_63_hexagon_filler.svg}
        \subcaption{\texttt{small\_005} start and intermediate positions}
        \label{subfig:ff1_small_005}
    \end{minipage}
    \hfill
    \begin{minipage}[t]{.46\textwidth}
        \centering
        \includesvg[scale=1.0]{graphics/ff2_small_005_10x10_90_63_hexagon_filler.svg}
        \subcaption{\texttt{small\_005} target and intermediate positions}
        \label{subfig:ff2_small_005}
    \end{minipage}\\
    \vspace{2em}
    \begin{minipage}[t]{.46\textwidth}
        \centering
        \includesvg[scale=1.0]{graphics/ff1_medium_005_30x30_90_407_hexagon_filler.svg}
        \subcaption{\texttt{medium\_005} start and intermediate positions}
        \label{subfig:ff1_medium_005}
    \end{minipage}
    \hfill
    \begin{minipage}[t]{.46\textwidth}
        \centering
        \includesvg[scale=1.0]{graphics/ff2_medium_005_30x30_90_407_hexagon_filler.svg}
        \subcaption{\texttt{medium\_005} target and intermediate positions}
        \label{subfig:ff2_medium_005}
    \end{minipage}
    \caption{A visualization of our initializations for the instances \texttt{small\_005}
    and \texttt{medium\_005}. The grey squares are obstacles.
    On the left (Figures~\ref{subfig:ff1_small_005} and~\ref{subfig:ff1_medium_005}), the colours of the filled boxes match the instances start positions to the intermediate positions.
    On the right, (Figures~\ref{subfig:ff2_small_005} and~\ref{subfig:ff2_medium_005}), the colours match the target positions and intermediate positions.}
    \label{fig:matching_result}
\end{figure}

\begin{figure}[ht]
    \centering
    \begin{minipage}[t]{.46\textwidth}
        \centering
        \includesvg[scale=1.0]{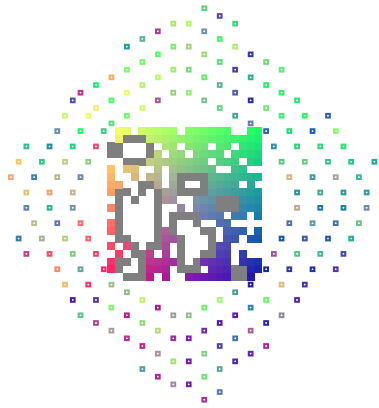}
        \subcaption{Octagon filler}
    \end{minipage}
    %\hfill
    \begin{minipage}[t]{.46\textwidth}
        \centering
        \includesvg[scale=1.0]{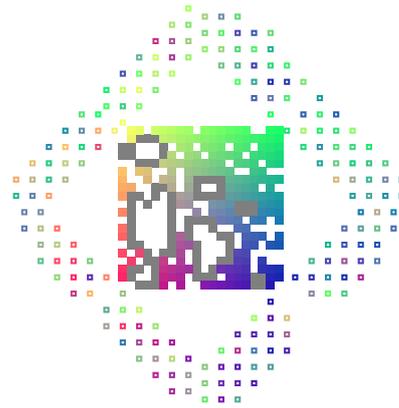}
        \subcaption{Diamond filler}
    \end{minipage}\\
    \vspace{2em}
    \begin{minipage}[t]{.46\textwidth}
        \centering
        \includesvg[scale=1.0]{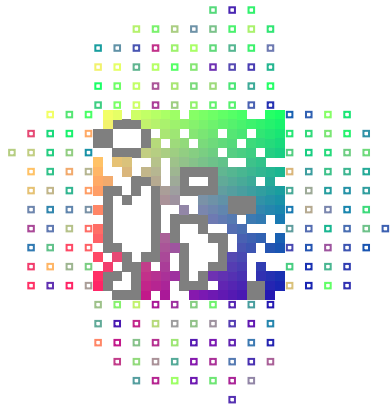}
        \subcaption{Quad rectangle filler}
    \end{minipage}
    %\hfill
    \begin{minipage}[t]{.46\textwidth}
        \centering
        \includesvg[scale=1.0]{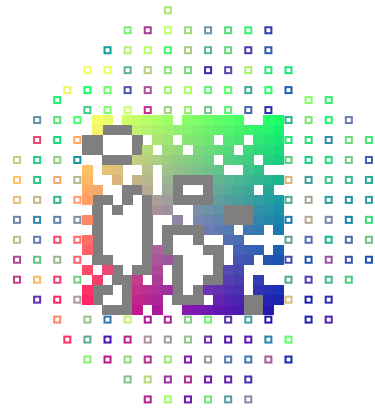}
        \subcaption{Rectangle filler}
        \label{subfig:rectfig}
    \end{minipage}
    \caption{Visualizations of the different methods we used to generate intermediate positions for the instance \texttt{small\_012}.
    This visualization only shows the matched potential intermediate positions,
    which is why the displayed intermediate positions in Figure~\ref{subfig:rectfig} do not resemble a rectangle.
    }
    \label{fig:filler}
\end{figure}

\begin{definition}
\label{def:depth}
Given a set of locations $H \subset \ZZ^2$, the depth value $D(v)$ for $v\in\ZZ^2$ is computed recursively:
\begin{itemize}
    \item If $v\in H$, then $D(v)=0$.
    \item Otherwise, $D(v)=1+\min_{u}D(u)$ where the $\min$ is over all non-obstacle squares adjacent to $v$.
\end{itemize}
Such values are uniquely defined and can easily be found via breadth-first search.
\end{definition}

The initialization operates in a few phases:
\begin{enumerate}
    \item Compute a set of intermediate positions $H$, derived from a superset of candidate positions which we call \emph{filler} shapes (see Figure~\ref{fig:filler}). These filler shapes are located outside of the bounding box of the robots' start positions, and arranged in a way such that no intermediate positions are adjacent.
    \item Compute a min-cost matching of the robots
    to their intermediate positions. The cost of a matching is the sum of distances from the robots' start and target positions to their matched intermediate position. To improve the initialization, we generate many more intermediate positions than robots.
    Figure~\ref{fig:matching_result} shows an example of the robots' start and target positions and the corresponding matched intermediate positions.
    \item For each coordinate, compute the depth of that coordinate using $H$ (see \Cref{def:depth}). Route the robots from their start position to their corresponding intermediate positions one at a time, in order according to decreasing depth values of their start positions. The routes here are constructed to avoid the paths of the robots routed before it. In this phase we route robots with starting locations of larger depth before smaller ones.
    \item Re-route the robots \emph{from their start positions} to their target positions' in decreasing order of their depth values. We start with the paths computed in the previous phase, and reroute each robot individually. We do this by removing their path and inserting a new path respecting the current paths of all robots.
\end{enumerate}

Although a natural alternative to step 4 is to simply route the robots from the intermediate locations to their target locations, it is instead much more efficient to re-route from the robots' starting locations.
Given the paths produced in the third step, a feasible set of paths is guaranteed to exist in the fourth step, since it's clear that there exists a set of paths from each robots' intermediate position to their target positions:
All that is needed is a path from the intermediate position to the target position that works after all other robots have finished moving.
Such a path exists since robots residing at higher depth values do not block the paths of robots at lower depth values.

\paragraph*{Initialization variations} 
Local search strategies are typically very sensitive to the initialization. 
We used a number of variations to generate a family of different initializations:
\begin{itemize}
    \item Using different filler shapes in step 1, or different costs for matching in step 3.
    \item Finding random shortest paths, or approximate shortest paths for steps 3 and 4.
    \item Swapping the robots' start and target locations.
\end{itemize}
The optimization procedure was run on the initializations that scored the highest.

\subsection{Solution optimization}
\label{sec:k-opt}
As previously described, the basic paradigm of our local optimization is to choose $k$ robots, and improve their paths while keeping the paths of all others fixed. This type of approach is often referred to as $k$-opt.
This approach has two distinct components: (1) the choice of the $k$ robots and (2) the improvements of their paths.

\paragraph*{Choosing the $k$ robots}
We experimented with a few types of weighted sampling to choose the $k$ robots:
\begin{itemize}
    \item \emph{Sampling by completion time.} Sample $k$ robots without replacement, where the sampling probability for each robot is proportional to their completion time (the time at which they make their last move).
    \item \emph{Sampling by closeness.} Sample the initial robot proportional to completion time, then sample $k-1$ other robots based on proximity to the initial robot's path. The proximity of two paths is the number of time steps for which the two paths are adjacent to each other.
    \item \emph{Sampling by constraints.} Sample the initial robot proportional to completion time, then find a minimum completion time path from the start to the target for the sampled robot with the relaxation that this robot is allowed to move through $k-1$ other robots. The robots that the path moves through forms the $k-1$ other robots in the sample. 
\end{itemize}
In our experiments, sampling by completion time and by closeness seemed to produce the best results. However, we note that team Shadoks was able to exploit a variant of sampling by constraints to win the MAX category. 

\paragraph*{Path optimization}
Given a sample of $k$ robots, one would like to jointly optimize the $k$ robots simultaneously.
However, this approach causes the state space to grow exponentially with the number of robots.
Furthermore, due to the size of the grid, the state space for path finding is already quite large to begin with.
Like other teams, our path finding was done in the grid-time graph where states are characterized by the positions of the robots at each time $t$.
For the largest instance in the data set, the size of this state space is already on the order of $10^5$. For these reasons, we were only able to scale to $k=2$ in our code with the approach of joint optimization.

Instead, our main insight was to \emph{approximate} the joint optimization by $k$ individual single path optimizations. 
Our inspiration was the analogy of sorting: given the $k$ robots, we route the robots one-by-one to their targets, where the $j$-th robot routed ($j \leq k$) respects the path of robots $1\ldots j-1$. Crucially, this avoids the exponential blow-up of the state space as we're routing each robot from start to target in succession.
The problem then reduces to finding a good ordering of the $k$
robots to reroute.
We had the most success by simply ordering the robots by decreasing completion time.
If an ordering was infeasible (e.g. due to some subset of the $k$ robots completely blocking off the path from the remaining robots) or not an improvement, it was discarded.

During the competition we found that there were advantages to relatively smaller values of $k$ for faster computation. For our approximate joint optimization, larger $k$ helped to get out of bad local optima. We iteratively used all values of $k$ between $1$ and $7$ during the challenge.

Since the main component of our algorithm is path-finding, we used a number of practical techniques to reduce its cost.
We use A* with a bucket priority queue, where the heuristic function was taken to be either the Manhattan distance or the shortest path distance in the graph with only obstacles without robots. 
To further speed up the search when we have an initial feasible solution, we limit the path-finding algorithm to search locally around the robot's original path,
by enforcing that no robot may deviate more than $R$ steps away from the set of positions forming their initial path for a fixed parameter $R$. Since the search is centred around the initial solution, the solution space is guaranteed to have a feasible solution for any value of $R$. 

The optimizations above are agnostic to the optimization objective and allowed for large gains following the initialization phase for both SUM and MAX.

\section{Results}
\subsection{Computational environment}
For the most part, our experiments were done on a desktop with a Core i7-2600. % Linux machine specs
In the last three weeks of the competition we used Stanford's Sherlock High-Performance Computing Cluster for SUM optimization.
The University of Waterloo's Multicore lab also generously donated some night-time compute during the last week of the competition in the form of two EPYC 7662s.
\subsection{Experimental results}
\label{sec:results}
\begin{figure}[ht]
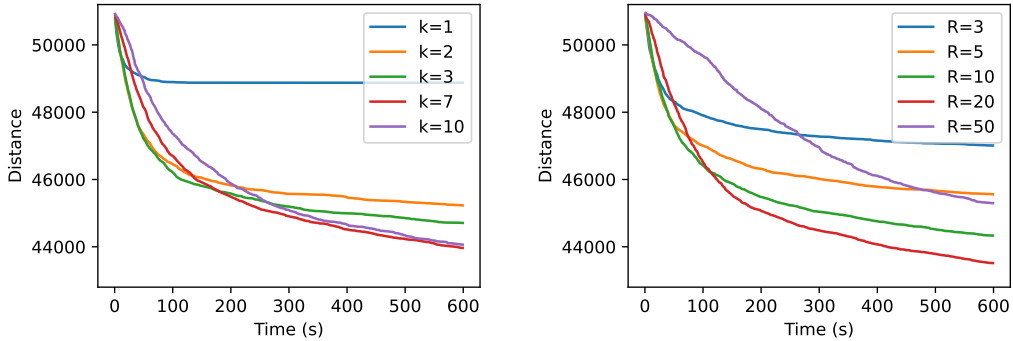

    \centering
    \begin{minipage}{0.49\textwidth}
        \centering
        \includesvg[width=\textwidth]{microbes_00004_50x50_50_1250_score_distance.svg}
        \subcaption{SUM optimization varying $k$ where $R=20$.}
        \label{subfig:distance_k_vary}
    \end{minipage}
    \begin{minipage}{0.49\textwidth}
        \centering
        \includesvg[width=\textwidth]{microbes_00004_50x50_50_1250_score_R_distance.svg}
        \subcaption{SUM optimization varying $R$ where $k=7$.}
        \label{subfig:distance_R_vary}
    \end{minipage}
    \caption{Varying $k$ and $R$ for SUM optimization on \texttt{microbes\_004}. Final SUM: 43437.}
    \label{fig:experimental_results_distance}
\end{figure}

\begin{figure}[ht]
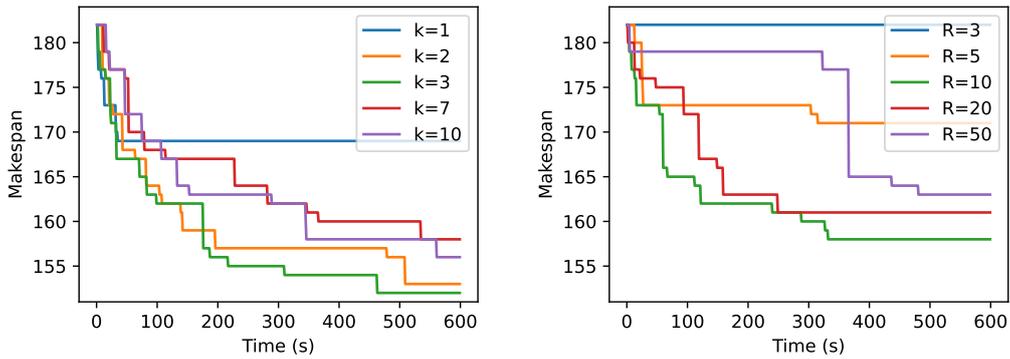

    \centering
    \begin{minipage}{0.49\textwidth}
        \centering
        \includesvg[width=\textwidth]{microbes_00004_50x50_50_1250_score.svg}
        \subcaption{MAX optimization varying $k$ where $R=20$.}
        \label{subfig:makespan_k_vary}
    \end{minipage}
    \begin{minipage}{0.49\textwidth}
        \centering
        \includesvg[width=\textwidth]{microbes_00004_50x50_50_1250_score_R.svg}
        \subcaption{MAX optimization varying $R$ where $k=7$.}
        \label{subfig:makespan_R_vary}
    \end{minipage}
    \caption{Varying $k$ and $R$ for MAX optimization on \texttt{microbes\_004}. Final MAX: 126.}
    \label{fig:experimental_results_distance_makespan}
\end{figure}

To empirically determine the effect of $k$ and $R$ had on the optimization, we ran our algorithm on the instance \texttt{microbes\_00004}.
As seen in Figure~\ref{subfig:distance_k_vary}, different values of $k$ had an effect on the rate of convergence and the value of the local minimum found by our strategy.
Experimentally, values of $k$ greater than $10$ did not improve our scores and also ran slower.
For MAX (in Figure~\ref{subfig:makespan_k_vary}), improvements were much more discrete. 

As we varied $R$, we had a similar trade-off between performance and runtime for each optimization step of our algorithm (Figures~\ref{subfig:distance_R_vary} and~\ref{subfig:makespan_R_vary}). Empirically, choosing $R$ to be around $20$ achieved the best balance for most of the instances.

\begin{figure}[!ht]
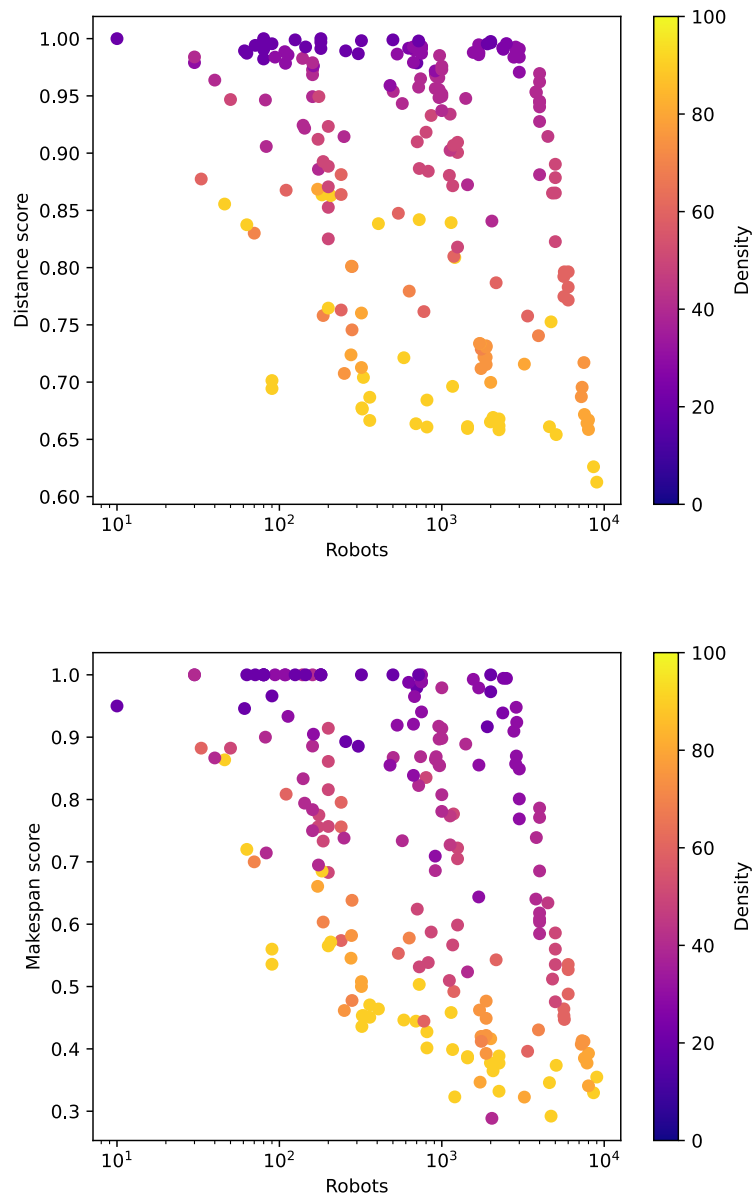

    \centering
    \includesvg[width=0.8\textwidth]{dist_scores.svg}
    \includesvg[width=0.8\textwidth]{span_scores.svg}
    \caption{Plots of $\frac{\text{score}}{\text{lower bound}}$ versus $n$ for both SUM and MAX. Each point corresponds to one instance. The density of the instance, as defined by the contest organizers, is indicated by colour.}
    \label{fig:experimental_results_density}
\end{figure}

Figure~\ref{fig:experimental_results_density} compares the solutions we computed during the competition to the trivial lower bound, which is given by the maximum shortest-path distance for SUM and the sum of all shortest-path distances for MAX.
There is a clear trend that score decreases as density increases (and as $n$ increases, to a lesser extent). This could be due to two factors: first, that the trivial lower bound becomes a worse approximation of the optimal score as these values increase; second, that our algorithm performs poorer on these instances, because more computational time is required and local changes to $k$ robots at a time are insufficient to traverse the solution space.

\bibliography{references}

\end{document}